\documentclass[hyper]{JHEP} 

\usepackage{epsfig}

\newcommand\fverb{\setbox\pippobox=\hbox\bgroup\verb}
\newcommand\fverbdo{\egroup\medskip\noindent%
			\fbox{\unhbox\pippobox}\ }
\newcommand\fverbit{\egroup\item[\fbox{\unhbox\pippobox}]}
\newbox\pippobox
\title{Some Remarks About Berkovits'Superstring Field
Theory}

\author{by J. Kluso\v{n}\\
	 Department of Theoretical Physics and Astrophysics\\
                   Faculty of Science, Masaryk University\\
Kotl\'{a}\v{r}sk\'{a} 2, 611 37, Brno\\
Czech Republic\\
	E-mail: \email{klu@physics.muni.cz}}

\preprint{\hepth{0105319}}

\abstract{In this short note we would like to discuss 
 general solutions of the  Berkovits superstring field theory,
in particular the string field action for fluctuation
around such a solution. We will find that fluctuations
obey the same equation of motion as the original
field with the new BRST operator. Then
we will argue that the superstring field theory
action for fluctuation field
 has the same form as the original one.}

\keywords{String field theory}

\begin{document}
\section{Introduction}\label{first}

In recent two years there was a great
interest in a  string field theory since
this is  only known nonperturbative
definition of the string theory allowing
off-shell computation of the tachyon
potential (For  review  and
extensive list of references, see
\cite{Ohmori}, some recent works considering
related problems are \cite{SenV1,Hata,Taylor,
Feng,Ohmori1,Feng1,SenV2,SenV3,Gross1,
Kawano,SenV4,Muki,David,Nakatsu,Michishita}. While 
  bosonic
string field  theory \cite{WittenSFT}
 based on the Chern-Simons-like
 action is very well understood 
  in superstring case
the situation is not such a clear. The straightforward
superstring extension 
of the bosonic string field theory
 \cite{WittenSFT1} suffers from an emergence
of contact divergences \cite{Wendt} that
forces us to include higher-contact
 terms when two picture changing
operators collide. (For discussion of 
this issue and other related problems we
again recommend to see very good review
\cite{Ohmori}.)
To overcome contact term divergences, a slightly
different formalism was proposed in
\cite{Thorn,Zubarev1,Zubarev2}. For recent 
discussion of this approach, see
\cite{Ohmori,Zubarev3}.

Another interesting approach to the problem
of the construction of superstring field theory
was given in Berkovits' works 
\cite{Berkovits2,Berkovits3} where Lorentz-
covariant action for  the Neveu-Schwarz
sector of open superstring field theory was
constructed.   While previous
string field theories are based on the
Chern-Simons-like actions, Berkovits formulation
is based on the Wess-Zummino-Witten-like
 action. Two major advantages
of Berkovits superstring field theory
 are 
that it does not require contact terms to
remove tree-level divergences
\cite{BerkovitsD} and that the calculation
of the tachyon potential  in case of non-BPS D-brane
\cite{Berkovits3,Sen1} 
gives a very good agreement with Sen's conjecture
 \cite{SenP} (For recent
review of Berkovits Neveu-Schwarz string
 field theory (NSFT), see \cite{BerkovitsR}.). 
The tachyon kink and lump solution was
also analysed (For recent work, see \cite{Ohmori1},
for list of references, see \cite{Ohmori}.) 
However in spite of these remarkable results
strongly supporting  validity and consistency of 
this approach, NSFT
is not without own problem since  it is not known
how to extend this action to the Ramond sector 
in ten dimensional Lorentz-invariant manner.
But it is commonly believed that this problem is
 technical nature only and we will succeed
in including of the Ramond sector into this theory
in the near future.

In this paper we will discuss NSFT from different
 point of view.
As is well known from the bosonic  case, 
when we expand the string field around any
 solution of the equation of motion,
 we find that the BRST operator has changed
while the string field action has the 
same structure as the original one (for a nice discussion 
of this approach in case of closed bosonic string field theory,
see, for example \cite{SenCS}, for recent discussion, see
 \cite{SenV1,SenV2,SenV3}). 
Then we can  ask  whether  similar
process occurs in NSFT as well. In fact, analysis
of this question could be helpful for extension
of recent results \cite{SenV1,SenV2,SenV3} to the
supersymmetric case and in the end could lead to
more fundamental formulation of string theory and
M theory. 

In order to study these solutions, we firstly
 rewrite NSFT action in a slightly different form that resembles
more similarity with the original Wess-Zumino-Witten (WZW)
model (For review, see \cite{Gawedski}).
 In this formulation  we will regard two operators
$Q_B,\eta_0$ as a part of generalised
 exterior derivative. Using this formulation we can
manipulate with the action as with the ordinary
WZW action using the language of differential
geometry so that we are able very easily to
find  variation of the action and hence an equation
of motion.

As a next step we will analyse 
 solution of the 
equation of motion arising from the variation  
of the action and  fluctuation field
around this solution. 
At first sight an action for fluctuation will appear to 
 be different then the original one thanks
to the  presence of the field we expand around
in it. In order to answer the question whether
the fluctuation fields obey the same equation of
motion  and hence whether
string field action is the same as the original one,
 we will not proceed in 
such a straightforward manner as in the bosonic
string field theory where the new  BRST
operator naturally emerges from
an action for fluctuating field. Rather we will
study the equation of motion for fluctuating
field and from it we will extract an information
about a new BRST operator in NSFT. We will show
on an example of the bosonic string field theory that
this is an equivalent approach for searching
of a new form of the BRST operator as the approach given
above. Then we will show that
in NSFT the BRST
operator is modified with the background solution 
exactly in
the same way as in the bosonic case.
We believe that this is a nontrivial result since
the form of the Berkovits superstring field theory
is completely different then the open bosonic string
theory. We also hope that our result could 
be helpful in the extension of the study
of the vacuum string field theory
 \cite{SenV1,SenV2,SenV3,Gross1,Kawano,SenV4}
to the supersymmetric case.

The plan of the paper is follows. In the
next section (\ref{second}) we briefly review 
NSFT theory. 

In section (\ref{third})
we rewrite  NSFT action in a slightly different form
that  resembles more similarity
with the ordinary WSW model. Using this formalism
we will be able to find a variation of the action
and hence an equation of motion.

In section (\ref{fourth}) we will discuss
  solution of the  equation of motion
arising from the variation of the action. Then
we will study the behaviour of any string
field around this new solution. 

In conclusion (\ref{fifth}) we will outline our
 results  and suggest
further extension of this work.

\section{Review of superstring field theory}
\label{second}
 
In this section we would like to review basic facts about
superstring field theory, for more details, see \cite{Ohmori,
Berkovits1,Berkovits2,Sen1}.
The general off-shell string field configuration
in the GSO(+) NS sector corresponding Grassmann
even open string vertex operator $\Phi$ of ghost number
$0$ and picture number $0$ in the combined
conformal field theory of a $c=15$ superconformal
matter system, and $b,c,\beta,\gamma$ ghost
system with $c=-15$. We can also express $\beta,
\gamma$ in terms of ghost fields
$\xi,\eta,\phi$
\begin{equation}
\beta= e^{-\phi}\partial \xi \ , 
\gamma=\eta e^{\phi} \ ,
\end{equation}
the ghost number $n_g$ and the picture number $n_p$
assignments are as follows
\begin{eqnarray}
b: \ n_g=-1,\ n_p=0 \ \ \ c: \ n_g=1, \ n_p=0 \ ; \nonumber \\
e^{q\phi}: \  n_g=0, \ n_p=q \ ; \nonumber \\
\xi: \ n_q=-1 , \ n_p=1 \ \ \ \eta: \ n_q=1, \ n_p=-1 \ . \nonumber \\
\end{eqnarray}
The BRST operator $Q_B$ is given
\begin{equation}
Q_B=\oint dz j(z)=\int dz
\left\{c(T_m+T_{\xi\eta}+T_{\phi})+c\partial cb+\eta e^{\phi} G_m
-\eta \partial \eta e^{2\phi}b\right\} \ ,
\end{equation}
where
\begin{equation}
T_{\xi\eta}=\partial \xi \eta, \
T_{\phi}=-\frac{1}{2}\partial \phi\partial\phi-\partial^2\phi \ ,
\end{equation}
$T_{m}$ is a matter stress tensor and $G_m$ is a matter superconformal 
generator. Throughout this paper we will be working in
units $\alpha'=1$.

The string field action is given \cite{Berkovits1,Berkovits2}
\begin{equation}\label{NSFTaction}
S=\frac{1}{2}\int\left(
(e^{-\Phi}Q_Be^{\Phi})(e^{-\Phi}\eta_0
e^{\Phi})-\int_0^1 dt
(e^{-t\Phi}\partial_te^{t\Phi})\left\{
(e^{-t\Phi}Q_Be^{t\Phi}),
(e^{-t\Phi}\eta_0e^{t\Phi})\right\}\right) \ ,
\end{equation}
where $\{A,B\}=AB+BA$ and $e^{-t\Phi}\partial_t
e^{t\Phi}=\Phi$. Here the products and
integral are defined by Witten's gluing prescription
of the string. The exponential of string field is
defined in the same manner $e^{\Phi}=
1+\Phi+\frac{1}{2}\Phi\star\Phi+\dots$. In the
following we will not explicitly write $\star$ symbol.
The basis properties of $Q_B,\eta_0$ which we
will need in our analysis (for more details, see
\cite{Ohmori} and reference therein) are
\begin{eqnarray}\label{ax}
Q_B^2=0, \ \eta_0^2=0, \ \{Q_B,\eta_0\}=0 \ ,
\nonumber \\
Q_B(\Phi_1\Phi_2)=Q_B(\Phi_1)\Phi_2+
\Phi_1Q_B(\Phi_2) , \ \nonumber \\
\eta_0(\Phi_1\Phi_2)=\eta_0(\Phi_1)\Phi_2+
\Phi_1\eta_0(\Phi_2) , \ \nonumber \\
\int Q_B(\dots)=0 \ , \int \eta_0(\dots)=0 \ , \nonumber \\
\end{eqnarray}
where $\Phi_1,\Phi_2$ are Grassmann even fields. 

In the next section we rewrite (\ref{NSFTaction}) in
the form that resembles more
 similarity with the original WZW model
and that allows us to find very easily variation of
 the action.

\section{Geometrical formulation of the superstring
field theory action}\label{third}
In this section we rewrite the NSFT action 
(\ref{NSFTaction}) in a geometrical formalism 
which will allow us to find very easily variation
of the action \cite{Berkovits1,Berkovits2} and
consequently  equation of motion.

 Let us define generalised exterior derivative
as follows
\begin{eqnarray}
dX=\partial_t X dt+ Q_B(X)dx^2+\eta_0(X)dx^3=
\partial_i X dx^i, \nonumber \\
  dx^2\wedge dx^3=dx^3\wedge dx^2 , \
dx^1\wedge dx^2=-dx^2\wedge dx^1, 
dx^1\wedge dx^3=-dx^3\wedge dx^1, \nonumber \\
dx^2\wedge dx^2=dx^3\wedge dx^3=0  \ , \nonumber \\
\end{eqnarray}
with any string field $X$.
Now we prove  nilpotence of this operator
$d^2X=0$.  Then we have
\begin{equation}
d^2(X)=d(\partial_i(X)dx^i)=
\partial_i(\partial_j X)dx^i\wedge dx^j=0 \ ,
\end{equation}
since for $i=1,j=2,3$ operators $Q_B,
\eta_0,\partial_t$ commute but
$dx^1\wedge dx^{2,3}=-dx^{2,3}\wedge dx^1$  and
for $i,j=2,3$ operators anticommute (note
$\{Q,\eta_0\}=0$) but $dx^1\wedge dx^2=dx^2\wedge dx^1$
 commute between themselves. We have also used
$Q_B^2=\partial_2(\partial_2)=0 , \
\eta_0^2=\partial_3(\partial_3)=0$.
Let us return to the  action
(\ref{NSFTaction}).
In the following we introduce a notation \cite{Sen1}
\begin{equation}
G=e^{\Phi}, \ \hat{G}=e^{t\Phi} \ .
\end{equation}
We start our analysis with the second term in (\ref{NSFTaction})
which we will write as
\begin{equation}\label{geWZW}
\frac{1}{3}\int \hat{G}^{-1} d\hat{G}\wedge
 \hat{G}^{-1} d\hat{G} 
\wedge \hat{G}^{-1} d\hat{G} \ ,
\end{equation}
where now the integral denotes integration over $t$
 and also abstract
Witten's string field integration \cite{WittenSFT}.
 We implicitly presume
that this  integration
is defined as
\begin{equation}
\int\int dt \left\{\dots\right\}=
\int \left\{\dots\right\}dx^1\wedge dx^2\wedge dx^3 \ .
\end{equation}
We define $\hat{G}(t)$ as a function of $t$
with the property that for $t=1, \hat{G}(1)=G, \ 
\hat{G}(0)=1$. We will show that  formulation (\ref{geWZW})
corresponds to  the second term in (\ref{NSFTaction})
\begin{eqnarray}
\frac{1}{3}\int \hat{G}^{-1}\partial_i\hat{G}
\hat{G}^{-1}\partial_j\hat{G}\hat{G}^{-1}
\partial_k\hat{G}
dx^i\wedge dx^j\wedge dx^k=\nonumber \\
=\frac{1}{3}\int \hat{G}^{-1}\partial_1 \hat{G}[
\hat{G}^{-1}\partial_2 \hat{G} \hat{G}^{-1}
\partial_3 \hat{G}+
\hat{G}^{-1}\partial_3 \hat{G} \hat{G}^{-1}\partial_2 \hat{G}]dx^1\wedge
dx^2\wedge dx^3+\nonumber \\+
\hat{G}^{-1}\partial_2 \hat{G}[
\hat{G}^{-1}\partial_3\hat{G}\hat{G}^{-1}\partial_1\hat{G}-
\hat{G}^{-1}\partial_1\hat{G}\hat{G}^{-1}
\partial_3\hat{G}]dx^1\wedge dx^2
\wedge dx^3+ \nonumber \\
+\hat{G}^{-1}\partial_3\hat{G}[\hat{G}^{-1}
\partial_2\hat{G}\hat{G}^{-1}\partial_1\hat{G}-
\hat{G}^{-1}\partial_1\hat{G}\hat{G}^{-1}\partial_2\hat{G}
]dx^1\wedge dx^2 
\wedge dx^3= \nonumber  \\
=\int \hat{G}^{-1}\partial_1 \hat{G}[\hat{G}^{-1}\partial_2 \hat{G}
 \hat{G}^{-1}
\partial_3 \hat{G}+
\hat{G}^{-1}\partial_3 \hat{G} \hat{G}^{-1}\partial_2 \hat{G}]dx^1\wedge
dx^2\wedge dx^3 \ , \nonumber \\ 
\end{eqnarray}
which we wanted to prove. In the previous expression we have used 
(We omit the symbol $dx^1\wedge dx^2\wedge dx^3$ which
does not change in the following manipulation)
\begin{eqnarray}
\int \hat{G}^{-1}\partial_2\hat{G}\hat{G}^{-1}
\partial_3\hat{G}\hat{G}^{-1}\partial_1\hat{G}=
\int \hat{G}^{-1}\partial_1 
\hat{G}\hat{G}^{-1}\partial_2\hat{G}\hat{G}^{-1}
\partial_3\hat{G} \ , \nonumber \\
-\int \hat{G}^{-1}\partial_2
\hat{G}\hat{G}^{-1}\partial_1 \hat{G}\hat{G}^{-1}\partial_3\hat{G}=
\int \hat{G}^{-1}\partial_1 \hat{G} 
\hat{G}^{-1}\partial_3\hat{G}\hat{G}^{-1}\partial_2\hat{G} \ ,
\nonumber \\
\int \hat{G}^{-1}\partial_3 \hat{G}\hat{G}^{-1}
\partial_2\hat{G}\hat{G}^{-1}\partial_1\hat{G}=
\int \hat{G}^{-1}\partial_1 \hat{G} \hat{G}^{-1}
\partial_3\hat{G} \hat{G}^{-1}
\partial_2 \hat{G} \ , \nonumber \\
-\int \hat{G}^{-1}\partial_3\hat{G}\hat{G}^{-1}
\partial_1\hat{G}\hat{G}^{-1}\partial_2\hat{G}=
\int \hat{G}^{-1}\partial_1\hat{G}\hat{G}^{-1}
\partial_2\hat{G}\hat{G}^{-1}\partial_3\hat{G} \ ,
\nonumber \\
\end{eqnarray}
which follows from   the definition of $Q_B,\eta_0$ operators. 
Using description (\ref{geWZW})  we can easily find the
variation of the WZW term
\begin{eqnarray}
\delta \frac{1}{3}\int (\hat{G}^{-1}d\hat{G})^3=
\int \delta \hat{G}^{-1} d\hat{G}
\wedge (\hat{G}^{-1} d\hat{G})^2+
\int \hat{G}^{-1}d\delta \hat{G} \wedge (
\hat{G}^{-1} d\hat{G})^2=\nonumber \\
=-\int \hat{G}^{-1}\delta \hat{G} 
\hat{G}^{-1} d\hat{G}\wedge (\hat{G}^{-1} d\hat{G})^2+
\int  d(\delta \hat{G} (\hat{G}^{-1} d\hat{G}
)^2\hat{G}^{-1})-\nonumber \\
-\int \delta \hat{G}
d ((\hat{G}^{-1} d\hat{G})^2 \hat{G}^{-1})=
\int d[\delta \hat{G} (\hat{G}^{-1}
d\hat{G})^2\hat{G}^{-1}] \ , \nonumber \\
\end{eqnarray}
where we have used
\begin{equation}
-\int \delta \hat{G} d((\hat{G}^{-1} d
\hat{G})^2\hat{G}^{-1})=
\int \hat{G}^{-1}\delta \hat{G} (\hat{G}^{-1} d\hat{G})^3 \ .
\end{equation}
We have also used the fact that  the exterior derivative acts
on various forms  as  
usual exterior derivative.
 More precisely, let $\omega=\omega_i dx^i,
\eta=\eta_i dx^i$ are one forms with Grassmann odd $\omega_{2,3}, \
\eta_{2,3}$ components. Then we have
\begin{eqnarray}
d (\omega \wedge \eta)=
\partial_k (\omega_i \eta_j)dx^k \wedge dx^i \wedge dx^j=
(\partial_k \omega_i dx^k\wedge dx^i )\eta_j\wedge dx^j-\nonumber \\
-\omega_i dx^i \wedge (\partial_k \eta_j dx^k \wedge dx^j)
=d\omega \wedge \eta-\omega \wedge d \eta \ , \nonumber \\
\end{eqnarray}
where the minus sign emerges either from the anticommutative
nature of $dx^1\wedge dx^{2,3}$ and the presence of the
ordinary derivative $\partial_1$
or through the anticommutative relation between the derivatives 
$\partial_{2,3}$ and $\omega_{2,3}, \eta_{2,3}$ 
and commutative nature of $dx^{2,3}$.
Then the variation of the WZW term is given (Remember, that in
our abstract interpretation the boundary of the "space"
 we integrate  over  is in the points $t=0,1$)
\begin{equation}
\left.\int \hat{G}^{-1}\delta \hat{G} 
(\hat{G}^{-1} d\hat{G})^2 \right |_{t=0}^{t=1} ,
\end{equation}
where the integration corresponds to Witten's
string field integration or in 
our notation to the integration  over abstract space spanned with
$x^{2,3}$ coordinates. The  fact that there is not 
 "boundary" of the space labelled with coordinates $x^2,x^3$
 can be also
seen from the definition  $\int Q_B(\dots)=
\int \eta_0(\dots)=0$.  Since we know that $\hat{G}(t)$ is a function
of $t$ with the property $\hat{G}(1)=G, \hat{G}(0)=1$ 
the variation of the action is equal to
\begin{equation}\label{varWZW1}
\int G^{-1}\delta G (G^{-1}Q_B(G)G^{-1}\eta_0(G)+
G^{-1}\eta_0(G)G^{-1}Q_B(G)) 
\end{equation}
and using the fact that the expression in the bracket can be written as
\begin{eqnarray}
-( \eta_0(G^{-1}Q_B(G))+Q_B(G^{-1}\eta_0(G)))=
-\eta_0(G^{-1})Q_B(G)-G^{-1}\eta_0(Q_B(G))-\nonumber \\
-Q_B(G^{-1})\eta_0(G)-G^{-1}Q_B(\eta_0(G))=
G^{-1}\eta_0(G)G^{-1}Q_B(G)+G^{-1}Q_B(G)
G^{-1}\eta_0(G) \  \nonumber \\
\end{eqnarray}
so that the variation of the WZW term is equal to
\begin{equation}\label{varWZW2}
\delta S_{WZW}=-\int G^{-1}\delta G\left(
\eta_0(G^{-1}Q_B(G))+Q_B(G^{-1}\eta_0(G))\right) \ .
\end{equation}

In the similar way we rewrite the first term in
(\ref{NSFTaction}). In order to do it we should
introduce the operation of  the Hodge dual $*$. Note
that the integration in the first term in 
(\ref{NSFTaction})  is performed on the space spanned
with $x^{2,3}$, so we can define the Hodge dual
operation for
this space  as follows
\begin{eqnarray}
* (dx^i)=\epsilon^i_jdx^j=g^{ik}\epsilon_{kj}dx^j, \
\epsilon_{23}=-\epsilon_{32}=1, \ 
 \nonumber \\
*(dx^2)=g^{23}\epsilon_{32}dx^2=-dx^2 , \
*(dx^3)=g^{32}\epsilon_{23}dx^3=dx^3 \nonumber \\ 
\end{eqnarray}
where $g^{ij}$ is an auxiliary metric with nonzero components $g^{23}=g^{32}=1$. 
Then we claim that the first term in 
(\ref{NSFTaction}) can be written as
\begin{equation}
\frac{1}{2}\int G^{-1}dG \wedge * G^{-1}dG \ ,
\end{equation}
since we have
\footnote{In this kinetic term the exterior derivative
does not contain $\partial_t$ derivative. Consequently
the integration $\int d\left\{\dots\right\}$ is equal to
zero thanks to the definition of $Q_B,\eta_0$.}
\begin{equation}
*G^{-1}dG=
G^{-1}\partial_2 G *(dx^2)+G^{-1}\partial_3G*(dx^3)
=-G^{-1}\partial_2 Gdx^2+G^{-1}\partial_3G dx^3 
\end{equation}
and then the kinetic term is equal to
\begin{eqnarray}
\frac{1}{2}\int (G^{-1}\partial_2 Gdx^2+G^{-1}\partial_3Gdx^3)
\wedge (-G^{-1}\partial_2Gdx^2+G^{-1}\partial_3Gdx^3)=
\nonumber \\ 
=\frac{1}{2}\int \left(
 G^{-1}\partial_2 G G^{-1} \partial_3G dx^2\wedge dx^3
-G^{-1}\partial_3 GG^{-1}\partial_2 G dx^2 \wedge dx^3\right)=
\nonumber \\ =
\int G^{-1}\partial_2 GG^{-1}\partial_3G dx^2\wedge dx^3
=\int G^{-1}Q_B(G) G^{-1}\eta_0(G) \ 
\nonumber \\
\end{eqnarray}
with the variation 
\begin{eqnarray}\label{varkin}
\delta \frac{1}{2}\int G^{-1}dG\wedge *G^{-1}dG=
\int \delta G^{-1}dG\wedge * G^{-1}dG+\int 
G^{-1}d\delta G \wedge * G^{-1}dG=
\nonumber \\
=-\int G^{-1}\delta G G^{-1}dG\wedge * G^{-1}
dG-\int \delta G d*(G^{-1}dG)G^{-1}-\int dG^{-1}\delta G
\wedge *G^{-1}dG=\nonumber \\
=-\int G^{-1}\delta G d\left( G^{-1}*dG\right)
=\int G^{-1}\delta G (\eta_0(G^{-1}Q_B(G))-Q_B(
G^{-1}\eta_0(G)))dx^2\wedge dx^3 \ . \nonumber \\
\end{eqnarray}
Collecting all previous results 
(\ref{varWZW1}), (\ref{varWZW2}),
(\ref{varkin}) we obtain
\begin{equation}
\delta S=-G^{-1}\delta G \left\{d\left(G^{-1}*dG\right)
+G^{-1}dG\wedge G^{-1}dG\right\}=
G^{-1}\delta G \eta_0(G^{-1}Q_B(G)) \ 
\end{equation}
so that the equation of motion has a form
\begin{equation}\label{eqm}
d\left(G^{-1}*dG\right)+G^{-1}dG\wedge G^{-1}dG=0
\Rightarrow \eta_0(G^{-1}Q_B(G))=0 \ ,
\end{equation}
with agreement with \cite{Berkovits1,Berkovits2,Sen1}.
It is clear that the upper result is not new but we
believe that our auxiliary "geometrical formulation" could
be helpful for better understanding of properties of the
NSFT action.

In the next section we use this result and
 we will study the  solution
of the equation of motion given above (\ref{eqm}).

\section{String field theory action for
fluctuation around solution of the equation
of motion}\label{fourth}

In this section we will discuss the
string field theory around any background configuration
which is a solution of (\ref{eqm}).
Since we will not perform any explicit
calculation we will again use abstract
Witten's formalism in string field theory
\cite{WittenSFT}. As usual we will not
write explicitly the string field theory
star product $\star$.  
Let us consider any string field $\Phi_0$, corresponding to
$G_0=e^{\Phi_0}$,  which is a solution of the equation 
of motion (\ref{eqm}) 
\begin{equation}\label{eqm1}
\eta_0(G_0^{-1}Q_B(G))=0 \ .
\end{equation}
Now we would like to study the fluctuation
around this  solution. For that reason
we write general string field containing 
fluctuation around this solution as
\begin{equation}
G=G_0h, \  h=e^{\phi}, \ G^{-1}=h^{-1}G^{-1}_0 \ . 
\end{equation}
To see  that this  field really describes
 fluctuations around  solution $G_0$ note
that for $\phi=0, G=G_0$. It is also clear that
any  string field in the form $e^{\Phi_0+\phi'}$ can be
always rewritten in the form  given above.

  Inserting this upper 
expression in (\ref{NSFTaction}) we obtain an action for
$\phi$
\begin{eqnarray}\label{actfluc}
S=\frac{1}{2}\int h^{-1}G_0^{-1}d(G_0 h) *
h^{-1}G_0^{-1}d(G_0h)-\nonumber \\-
\frac{1}{3}\int 
\hat{h}^{-1}\hat{G}_0^{-1}d(\hat{h}\hat{G_0})\wedge
\hat{h}^{-1}\hat{G}_0^{-1}d(\hat{h}\hat{G}_0)
\wedge \hat{h}^{-1}\hat{G}_0^{-1}d(\hat{h}\hat{G}_0) \ .
\nonumber \\
\end{eqnarray}
Now we would like to ask the question what form
of the equation of motion obeys shifted field $h=e^{\phi}$.
 We will show
on an example of the bosonic string field theory that
this is a meaningful approach for extracting an information
about  new BRST operator. In particular, it is well known
that the WZW term cannot be expressed in any closed
form (For review, see \cite{Gawedski}.)
 as an integral over two dimensional space spanned with
$x^2,x^3$ so it would be difficult to extract 
 the form of the new BRST operator
directly from
(\ref{actfluc}). On
the other hand, as we have seen in the previous section,
 the variation of the action can be written 
as an integral over $x^{2,3}$ so that it seems 
to be more efficient to gain information 
about new $Q_B,\eta_0$ from the equation of motion.  

In order to obtain this equation we will vary
(\ref{actfluc}) 
with respect to $G=G_0h, \delta G=G_0\delta h, 
\delta h= G_0^{-1}\delta G$ so we get
\begin{equation}
d (h^{-1}G_0^{-1} * d(G_0 h))+
G_0^{-1} h^{-1} d(G_0 h)\wedge 
h^{-1} G_0^{-1} d(G_0 h)=0 \ ,
\end{equation}
or equivalently
\begin{equation}
\eta_0(h^{-1}G_0^{-1}Q_B(G_0h))=
\eta_0(h^{-1}Q_B(h)+h^{-1}G_0^{-1}
Q_B(G_0)h)=0 \ .
\end{equation}
We can add to the upper expression
the term 
\begin{equation}
-\eta_0(G_0^{-1}Q_B(G_0))=
-\eta_0(h^{-1}h(-1)^{h} G_0^{-1}Q_B(G_0))
\end{equation}
 that is
equal to zero  according
to (\ref{eqm1}). 
We have chosen minus sign in the upper expression
since possible new BRST operator 
 should   obey all string
field theory axioms (\ref{ax}) and as we will
see minus sign in upper expression
 is crucial for it. Then we get
the final result
\begin{eqnarray}
\eta_0 (h^{-1}Q_B(h)+h^{-1}Ah-h^{-1}hA))=
\eta_0(h^{-1}\tilde{Q}_B(h))=0 \ , \nonumber \\
\tilde{Q}_B(X)=Q_B(X)+AX-(-1)^XXA \ , 
\forall X, \ A=e^{-\Phi_0}Q_B(e^{\Phi_0}) 
 \ . \nonumber \\
\end{eqnarray}
Upper expression suggests that the BRST operator
has changed through the  solution of the
equation of motion. Note that $A$ has a correct
properties to be added to the BRST operator since has a
ghost number one and picture number zero. We
also see that $\eta_0$ does not change 
which  from our point of view seems to
be natural result since $\eta_0$ is related to
the  ghost fields so that
it should not depend on any  background
configuration.

Let us compare this situation with Witten's 
bosonic open string field  theory where
 the string  field 
action is \cite{WittenSFT}
\begin{equation}\label{WCS}
S=\frac{1}{2}\int \Psi Q_B\Psi+\frac{1}{3}
\int \Psi \Psi\Psi \ .
\end{equation}
Let $\Psi_0$ be  solution of the equation of motion
\begin{equation}
Q_B\Psi_0+\Psi_0\Psi_0=0
\end{equation}
and let any string field $\Psi$ containing fluctuation
around this solution has a  form $\Psi=\Psi_0+\psi$. When we insert this
string field into (\ref{WCS}) and perform variation
we get an equation of motion for $\psi$ 
\begin{equation}
Q_B\psi+
\psi \Psi_0+\Psi_0\psi +\psi \psi+(Q_B\Psi_0+\Psi_0\Psi_0)
=0 \Rightarrow
Q_B'\psi+\psi\psi=0 \ ,
\end{equation}
where the new BRST operator $Q'_B$ is defined as 
\begin{equation}
Q'_B(X)=Q_BX+\Psi_0 X-(-1)^XX\Psi_0 \ ,
\end{equation}
where $X$ labels grading of  string
field $X$. 
In other words, the BRST operator
$Q_B'$ acting on  a field $\psi$ 
is not the same as the original one which
is certainly natural result since the field $\psi$
propagates around different background
then original string field $\Psi$.
We also see that the bosonic string field
action is the same as the original one as
it should be since its form should not
depend on any particular background CFT.
We must also say that we could obtain
the form of the new BRST operator $Q'_B$ directly
from the action (\ref{WCS}) without writing
the equation of motion for $\phi$ as is
well known for a long time.
 We have chosen the second
approach which gives an equivalent result
in order to see direct relation with the 
analysis performed in case of NSFT.

From this example we see  a striking 
similarity with Berkovits string field theory
even if it seems to be  difficult to
see that the action (\ref{actfluc}) has the
same form as the original one (\ref{NSFTaction}),
however with  the different BRST operator $\tilde{Q}_B$.
We  determine this action 
using the fact that the equation of motion for $\phi$
are the same as the equation of motion
obtained from (\ref{NSFTaction}) so it is natural
that it arises from an action with the same
form as (\ref{NSFTaction}) and with the BRST
operator $\tilde{Q}_B$. Hence the
action for fluctuation is
\begin{equation}\label{actfluc1}
S(\phi)=\frac{1}{2}\int e^{-\phi}\tilde{Q}_B(e^{\phi})
e^{-\phi}\eta_0(e^{\phi})-\int_0^1 dt 
e^{-t\phi}\partial_te^{t\phi}\left\{
e^{-t\phi}\tilde{Q}_B(e^{t\phi}),e^{-t\phi}\eta_0(e^{t\phi})\right\}
 \ .
\end{equation}
Note that in the upper   expression we have shifted the
action so that does not contain constant part $S(G_0)$.

To finish our analysis we should prove that the 
BRST operator $\tilde{Q}_B$ obeys all string field
theory axioms. We firstly prove  its nilpotence
\begin{eqnarray}
\tilde{Q}_B^2(X)=\tilde{Q}_B(Q_B(X)+AX-(-1)^XXB)=
Q_B(Q_B(X)+AX-(-1)^XXA)+\nonumber \\
+A(Q_B(X)+AX-(-1)^XXA)-
(-1)^{X+1}(Q_B(X)+AX-(-1)^XXA)A
=\nonumber \\=
(Q_B(A)+AA)X-X(Q_B(A)+AA)=0 \ , \nonumber \\
\end{eqnarray}
where we have used
\begin{equation}\label{QA}
Q_B(A)=Q_B(G_0^{-1}Q_B(G_0))=
-G_0^{-1}Q_B(G_0)G^{-1}_0Q_B(G_0)=-AA \ .
\end{equation}
We can also see that
\begin{eqnarray}
\tilde{Q}_B(XY)=Q_B(XY)+AXY-(-1)^{X+Y}XYA=\nonumber \\=
(Q_B(X)+AX-(-1)^XXA)Y+(-1)^XX(Q_B(Y)+\nonumber \\
+AY-(-1)^YYA)=\tilde{Q}_B(X)Y+(-1)^X\tilde{Q}_B(Y)
\nonumber \\
\end{eqnarray}
and
\begin{equation}
\{\eta_0,\tilde{Q}_B\}(X)=\eta_0(AX-(-1)^XXA)+
A\eta_0(X)-(-1)^{X+1}\eta_0(X)A=0 \ ,
\end{equation}
since $\eta_0(A)=0$ as follows from (\ref{eqm1})
and consequently $\eta_0(AX)=-A\eta_0(X)$. 
And it is also easy to see that
\begin{equation}
\int \tilde{Q}_B(X)=0 \ .
\end{equation}
We must say few words about this result. 
It is interesting that the nilpotence of the new
BRST operator $\tilde{Q}_B$ does not depend on
the fact that $G_0$ is a solution of the equation
of motion which is a difference with the bosonic
string field theory where the nilpotence of the new
BRST operator depends on  the fact that $\Psi_0$
is the solution of the string field theory equation of motion.
On the other hand we have seen that the requirement, that
the  anticommutator $\{\tilde{Q}_B,\eta_0\}$
must be equal to zero, can be obeyed only when
 $G_0$ solves
(\ref{eqm1}). In other words, the requirement that $\Phi_0$
is a solution of the equation of motion is crucial
in NSFT  as well, however from different reason then
in the bosonic string field theory.

\section{Conclusion}\label{fifth}
The main goal of this paper was to
study the fluctuation around any
solution of the Berkovits superstring
 field theory  \cite{Berkovits1,Berkovits2,BerkovitsR}.
For that reason we have rewritten
the action into slightly different form
which allows us to find very easily its
variation and hence an equation of motion.
Even if this formulation is certainly not
new, in the context of the superstring field
theory could give better sight how the
variation and hence string field theory
equation of motion arises from (\ref{NSFTaction}).

Then we have studied solution of the
equation of motion obtained from the NSFT
 action. We have been mainly concerned with
the question of the form of the action for
fluctuation around this  solution.
We have argued that thanks  to the nontrivial
form of the NSFT action it is difficult to
see directly from it a form of  a new BRST operator.
Then we have  shown on an example
of the bosonic string field theory that we
can obtain an information about the BRST operator
from an equation of motion. When we have
applied this method to the NSFT case we
have seen that the BRST operator changes in
the similar way as in the bosonic case
as we could expect. Then we have shown
that the new BRST operator obeys all axioms
of the NSFT. 

We believe that the  results presented in this
paper are nontrivial. In particular, we hope
that they could be helpful in the extension
of the recent interesting works  considering
vacuum string field theory \cite{SenV1,SenV2,SenV3,
Gross1,SenV4} to the supersymmetric case that
hopefully in the end could give a new insight into 
the structure of the superstring theory.  
\\
\\ 
{\bf Acknowledgement}
We would like to thank Rikard von Unge for
very useful discussions and especially
 Ashoke Sen for  correspondence about this
subject. We would like also thank Lubo\v{s}
Motl and D. M. Belov for critical comments of
the first version of this paper. 
\\
\\

\end{document}